\documentstyle[12pt]{article}
\title{Note on the Jarzynski Equality}
\author{E. G. D. Cohen\\
and\\
David Mauzerall\\
The Rockefeller University\\
New York, NY 10021}

\begin{document}
\maketitle

\begin{abstract}
The Jarzynski Equality relates the free energy difference between
two equilibrium states of a system to the average of the work over
all irreversible paths to go from one state to the other.  We
claim that the derivation of this equality is flawed, introducing
an ad hoc and unjustified weighting factor which handles
improperly the heat exchange with a heat bath. Therefore Liphardt
et al's experiment cannot be viewed as a confirmation of this
equality, although the numerical deviations between the two are
small. However, the Jarzynski Equality may well be a useful
approximation, e.g. in measurements on single molecules in
solution.
\end{abstract}

\baselineskip=1.5\baselineskip
\section{Introduction}

\hspace{.25in} Ever since Jarzynski derived a remarkable equality,
the Jarzynski Equality$^{[1,2]}$, (JE), his results have been
widely used: duplicated theoretically and tested experimentally.
It is impossible to do justice to all publications that have
appeared in connection with the JE: a survey article would be
needed for that and, in fact, one covering part of the material
has been written by Ritort$^{[3]}$. However, in this paper we want
to scrutinize the original general derivation of the JE$^{[1,2]}$
and discuss briefly a crucial experiment by Liphardt et al$^{[4]}$
which was designed to check the JE.  The communities accepting
the JE consist overwhelmingly of chemists and biophysicists to
which also one of us (DM) belongs, while the physicists (to which
EGDC belongs) have divided opinions.  We hope that this paper may
clarify and unify the various aspects of the JE.

The main point of this paper is to argue that the JE is not an
equality in any mathematical sense, but can be a useful
approximate equality in certain important fields, like, e.g. the
study of single molecules in solution.

The JE equates the difference between the free energy of two
equilibrium states $A$ and $B$ of a system in contact with a heat
bath, with an average over the irreversible work done over all
paths from $A$ to $B$. This work is done in general by a set of
external force centers, which can be characterized by a set of
time dependent parameters $\{\lambda_j(t)\}$ in the Hamiltonian of
the system: $H \equiv H(\{p_i(t)\}, \{q_i(t)\}; \{\lambda_j(t)\})
= H(\Gamma(t); \lambda(t))$. Here $\Gamma(t) \equiv
\{p_i(t)\},\{q_i(t)\}$, defines the microscopic state of the
system in its phase space, by giving the values of the momenta
$\{p_i(t)\}$ and coordinates $\{q_i(t)\}$ respectively, of the $i
= 1,2,..., N$ particles of the system at time $t$.  For simplicity
we will restrict ourselves to only one parameter $\lambda(t)$, so
that $j = 1$. One is interested in the Helmholtz free energy
difference $\Delta F=F_B-F_A$ between the initial and final
equilibrium states $A$ and $B$. By considering the work $W$ done
on the system when the external forces bring the system from state
$A$ to state $B$ along all possible {\it{irreversible}} paths,
with weights $e^{-\beta W_{irr}}$,  the JE reads:
\begin{equation}
<e^{-\beta W_{irr}}> = e^{-\beta \Delta F}
\end{equation}
where $\beta = 1/k_BT$, with $T$ the temperature of the heat bath
and $k_B$ Boltzmann's constant.  Since in (1) the average $< \; >$
is over the work $W_{irr}$ done over all possible irreversible
paths in phase space from $A$ to $B$, the derivation of the left
hand side of Eq.(1) proceeds microscopically via statistical
mechanics.

We will consider here the case that the system is for all times $t
\geq 0$ in thermal contact with a heat bath. The heat exchange
with the heat bath can be reversible or irreversible, depending on
whether it takes place with the internal temperature $T_I$ of the
system equal or not equal to the external temperature $T_E$ of the
heat bath surrounding the system, respectively. In the former case
the system will always be in a (canonical) equilibrium state
during the work process. As is explained in textbooks of
Thermodynamics, a slow or fast work process does not guarantee at
all its reversibility or its irreversibility, only the equality of
internal and external parameters does.

At $t=0$ the system in state $A$ is coupled to a heat bath of
temperature $T$, so that its temperature is also $T$.  It has then
a canonical distribution given by$^{[1]}$:
\begin{equation}
f(\Gamma(0),0) = \frac{1}{Z_0} \; e^{-\beta  H (\Gamma(0),
\lambda(0))}
\end{equation}
Here $Z_0 = Z_A$ is the canonical partition function at $t=0$,
where $\lambda = \lambda (0) = \lambda_A$. We assume that work is
done on the system during a total time $\tau$, so that $0 \leq t
\leq \tau$, while $\lambda(t)$ goes from $\lambda(0) = \lambda_A$
to $\lambda(\tau) = \lambda_B$, the final value of $\lambda$ which
corresponds to the equilibrium state $B$. The (mechanical) work
$W$ done on the system in an initial micro phase $\Gamma(0)$ over
a time $t$ will be given by:
\begin{equation}
W_t \equiv W(\Gamma,t) = \int^t_0 dt' \frac{\partial H (\Gamma
(t'), \lambda(t'))}{\partial \lambda} \; \; {\dot{\lambda}}(t')
\end{equation}
where ${\dot{\lambda}} (t') = d \lambda(t')/dt'$.

We will assume here for simplicity that the rate of change of
$\lambda$, i.e. ${\dot{\lambda}}$, is constant.

However, if the change of $\lambda (t)$ with $t$ is such that
irreversible processes are induced in the system driving it
possibly far from equilibrium, then, in order to reach an
equilibrium final state $B, \lambda (t)$ must be such that the
equilibrium state $B$ can actually be reached. During this
transition, heat exchange will take place (and possibly work will
be done). This will depend on the procedure of varying $\lambda$
with time and on the nature of the induced irreversible processes.
The change $\Delta F$ is computed by Jarzynski considering only
the mechanical work in going from $A$ to $B$.

\section{JE for thermostatted system}

\hspace{.25in} Since the system is in constant contact with a heat
bath, it becomes important whether the (constant) rate of change
${\dot{\lambda}}$ of $\lambda$ allows thermal equilibrium to be
maintained at all times between the system and the heat bath. If
it does, the process is reversible{\footnote{This is usually
formulated by requiring that the external changes induced in the
system are slow when compared to the internal relaxation times
needed to return to thermal equilibrium once a change of the
system (e.g., its volume) has taken place. In that case the
temperature of the system and that of the heat bath will always
remain the same during a measurement.}}, if it does not then the
process is irreversible, since the internal and external variables
will then not always be equal, or may not even be definable in the
system.

In addition, the reversibility can be characterized in a different
way, viz., the strength of the coupling of the system to the heat
reservoir or the rate of heat transfer, which we will call
${\dot{c}}$. It is this quantity together with the work rate
${\dot{w}} = {\dot{\lambda}} \frac{\partial H}{\partial \lambda}$,
which will determine whether the system will remain in thermal
equilibrium at all times i.e., whether a reversible or
irreversible process takes place. We can consider now several
cases.

\begin{itemize}
\item[a)] The rate of work done on the system ${\dot{w}}$ is very
small $({\dot{\lambda}} \approx 0)$ and the coupling ${\dot{c}}
\approx 0$ also. Then depending on the ratio
${\dot{w}}/{\dot{c}}$, thermal equilibrium can be maintained
between the system and the heat bath, at all times $0\leq t \leq
\tau$ when ${\dot{w}} < < {\dot{c}}$, i.e., there will be plenty
of time for the system to exchange heat with the reservoir, so
that the disturbance of the system due to the work done on the
system and which brings it out of thermal equilibrium, can be
readjusted for all $0 \leq t \leq \tau$ so that the system is
always in thermal equilibrium with the heat bath at temperature
$T$. If this obtains during the entire time $\tau$ that work is
done on the system then the work process is isothermal and
reversible.

\item[b)] If the ${\dot{w}}$ is very small, but the coupling
${\dot{c}}$ is not, then the work process will also be isothermal
and reversible.

\item[c)] If, on the other hand, ${\dot{w}}$ is very large and the
coupling ${\dot{c}}\approx 0$, then there is no way to maintain
thermal equilibrium during the work time $\tau$ and the system is
thermodynamically in a non-equilibrium state.

 \item[d)] However, if ${\dot{w}}$ is very large, then ${\dot{c}}$ has to
be sufficiently larger to maintain thermal equilibrium, if at all
possible, and in general one cannot assume that this is so.
\end{itemize}

\section{Critique}

The above considerations are relevant because, during the
evolution of an initial phase $\Gamma(0)$ of the system, it will
not only be subject to the mechanical work done by the external
forces via the Hamiltonian $H$ on the system alone, but also to
the simultaneous energy exchange with the heat bath. Therefore, to
know the mechanical work and the heat separately along a phase
space path, one has, in principle, to know the microscopic state
of both the system, the heat bath and their coupling, so as to
know whether, and if so, how much and in what direction, heat
exchange between the system and the heat bath has taken
place.{\footnote{For a system in contact with a heat bath the work
done on the system is not just the mechanical work, which will
change the system's internal energy $(dE)$, but also work
associated with the heat (energy) exchange with the bath $(TdS)$,
can already be seen in the isothermal reversible case, where $dW =
dE - dQ = dE - TdS = d(E-TS) = dF^{[5]}$.}}

Consequently the introduction in phase space of the weight
$e^{-\beta W_{irr}}$, for every irreversible (stochastic) path,
where the inverse temperature $\beta$ of the heat bath is used in
the weight for every $W_{irr}$, does not seem to make physical
sense. In fact, even if the mechanical work $W_{irr}$ itself could
be precisely determined, the canonical weight $e^{-\beta W_{irr}}$
associated with it, especially when no internal (system)
temperature is known or can even be defined - a possibility of
being far from equilibrium, which Jarzynski notes himself$^{[2]}$
- seems completely arbitrary and the use of the heat bath
temperature $1/\beta$ - the only known temperature available -
without foundation. As a consequence, an average over the
microscopic $e^{-\beta W_{irr}}$, as carried out in Eq.(1), does
not seem physically meaningful unless they are very close to
reversible ones and provide then a good approximation to those.
Therefore the JE is correct but trivial if $W=W_{rev}$ and it
seems that the use of $W_{irr}$ instead of $W_{rev}$ is, in
general, unfounded. A further discussion of this point can be
found in points of section 5.

In the above we argued that the heat exchange has not been
properly taken into account and an essentially only mechanical
theory has been used to derive Eq.(1), while, however, also
non-mechanical work e.g. thermal expansion due to heat energy has
to be considered.

A striking example of this is Jarzynski's remark$^{[2]}$ that it
would suffice to let the system evolve from parameter values
$\lambda_A$ to $\lambda_{B^*} = \lambda_B$ regardless of whether
$B^*$ is the equilibrium state $B$, whose free energy difference
$\Delta F$ with state $A$ one wants. As Jarzynski clearly states,
the {\it{non-equilibrium}} state $B^*$ at $\lambda_{B^*} =
\lambda_B$, reached {\it{before}} the equilibrium state $B$, can
easily be transformed into the desired equilibrium state $B$,
while the amount of (mechanical) work done to get from $\lambda_A$
to $\lambda_B$ remains the same. The only way, however, to go from
a non-equilibrium state $B^*$ with $\lambda_B$ to an equilibrium
state at $B$, with no more work done, is through the contact with
the heat bath at temperature $T$. Since the free energy $F$ will
be a minimum in equilibrium, the non-equilibrium $\lambda_{B^*} =
\lambda_B$ state will, {\it{only if properly chosen}}, indeed go
to the equilibrium state $B$, accompanied, however, with an
exchange of an unknown amount of heat and possibly work ,which
depend on $B^*$, all of which will change the free energy.

So far we have discussed the general physical theory of Jarzynski
as found in refs.[1,2]. This theory is supposed to hold for
systems of any size. As Jarzynski remarks$^{[2]}$ to obtain
observable effects the systems have to be small, since otherwise
the fluctuations of the work values become too small to be
observed. In this connection measurements on single molecules in
aqueous solutions seem very appropriate as a check on the JE.

\section{Experiment}

Liphardt et al$^{[4,6]}$ have carried out such an experiment on
the free energy difference (in their case the Gibbs free energy
$\Delta G$ instead of $\Delta F$) between the unfolded and the
folded conformations of a single P5abc RNA molecule, suspended
between two handles, in an aqueous salt solution to check the JE.
The experiments were carried out carefully, by stretching a single
RNA molecule many times, between the folded and unfolded
conformations $A$ and $B$. In this case the irreversibility of the
procedure manifests itself in hysteresis curves associated with a
cycle $A$ (folded) $\rightarrow B$ (unfolded) $\rightarrow A$
(folded). A number of constant stretching (switching) rates of 2 -
5 pN/s (slow) and 34 pN/s and 52 pN/s (fast), (the
${\dot{\lambda}}$ above), between folded and unfolded states were
applied and histograms were made of the work done versus the
extension of the molecule. To obtain statistics, seven independent
sets of data were collected for seven different RNA molecules with
a slow step between two fast steps and about 40
unfolding-refolding cycles per molecule were performed i.e. about
300 independent measurements were made. Results for the average
work differences for these three switching rates relative to the
reversible work were plotted in bins of about 0.7 $k_BT$. Plotting
for each switching rate the most probable value (i.e. the maximum
of the Gaussian fits to their histograms) of the work done on each
molecule, the fast extension (switching) rates produced work
values  $\approx 2-3 k_BT$, above the expected thermodynamic free
energy difference of about 60 $k_BT$, because of irreversible
contributions (apart from measurement errors) to the work, in
agreement with the Second Law of Thermodynamics. On the other
hand,  an average for each switching rate, using the JE Eq.(1)
gave values for the expected free energy difference within their
estimated error.

We note that the deviations from equilibrium for even the fastest
switching time, are in fact only 5\% (or about $3k_BT$), which
would not be unusual for such measurements. In addition, the
thermal equilibration time is of the order of picoseconds, while
we estimate the structural equilibration time to be of the order
of milliseconds$^{[7]}$ and the experimental switching times of
the order of seconds (slow) to 0.1 seconds (fast). Therefore since
${\dot{w}} < {\dot{c}}$, i.e., essentially reversible isothermal
measurements (cases a) and b) above) were performed. Considering
the above mentioned time scales, the entire experiment could well
be very close to an isothermal one for all switching rates used,
so that the JE equality gives results very close to those of using
only reversible paths in phase space.{\footnote{The remark by
Ritort$^{[3]}$ that a system may be far from equilibrium even when
close to an isothermal process because of its small size does not
seem to be correct. The small size will allow the fluctuations to
be measurable. However, they have no bearing on the fast change in
the control parameter, needed for deviations far from
equilibrium.}}

A few additional remarks on this experiment follow.

\noindent 1. The Liphardt et al. data appear to fit better to the
JE for large stretching than to the second cumulant or the
``Fluctuation Dissipation Relation (FD)'': $<W_{irr}> = W_{rev} +
\beta \sigma^2/2$, where $\sigma^2=<W_{irr}^2> - <W_{irr}>^2$ is
the variance of $W_{irr}${\footnote{We note that their FD is
derived from the JE, thereby assuming its validity, which is
supposed to be checked in the experiment.  This manifests itself
in the appearance of only work contributions to $\sigma^2$, while
$\Delta F$ should also contain, in principle, entropic
contributions.}} with $W_{irr} = W - W_{rev} = W_d$, the
dissipative work. This, in spite of the fact the system is not far
from equilibrium.

We note that in the table S1 of the supplementary material their
$<W_d>$ (scaled with $\beta$) for various extensions and switching
rates are rather close to $\sigma$, except $<W_d>$ in the
reversible case (very slow switch) where $<W_d>$ is assumed to be
zero. If one computes $\sigma^2/2$, using the $\sigma$ in the
second column of $S$1, the resulting $<W_d>$ is too large. If, on
the other hand, one corrects the $\sigma$'s in the second column
for the measurement error by a simple subtraction of the
$\sigma$'s at $<W_d>=0$, the estimates for $<W_d>$ are too small.
This bracketing of $<W_d>$ indicates that the data are compatible
with $<W_d>=\sigma^2/2$.

2. We will now derive the JE under three assumptions:

1) The $W_{irr}$ all have the weight $e^{-\beta W_{irr}}$, where
$\beta$ is the temperature of the heat bath, not the system;

2) a Gaussian assumption is made for the measured non-equilibrium
distribution functions;

3) a FD of $<W_d> = \frac{\beta \sigma^2}{2}$ has to be used.

Assuming that the distribution functions for $W_d$ are all
Gaussian and using those to fit the data, the left hand side of
the JE in eq.(1) can be written as:
\begin{eqnarray}
<e^{-\beta W_d}> & = & \int d\beta W_d \exp-[\beta W_d - (\beta
W_d - \beta^2 \sigma^2/2)^2/2\sigma^2] \\ \nonumber & \cdot &
\{\int d\beta W_d \exp - [\beta W_d - \beta^2
\sigma^2/2)^2/2\sigma^2]\}^{-1}
\end{eqnarray}

Carrying out the Gaussian integrals leads then to:\\
$<e^{-\beta W_d}>$ = 1 or $<e^{-\beta W_{irr}}> = e^{-\beta
W_{rev}} = e^{-\beta \Delta F}$ i.e. the JE. This result depends
crucially on the validity of $<W_d> = \frac{\beta \sigma^2}{2}$.\\
We emphasize that two of these three assumptions, although
approximately correct, are not justified, as explained above,
since the first two are not satisfied in non equilibrium systems.\\
3. We note that the often quoted claim $^{[4,11,12]}$ that the
origin of the correctness of the JE is because of the
over-weighting of the negative dissipative work balanced by the
under weighting the positive dissipative work has no known basis.
That the JE works here, in spite of the fact that it is based on
unfounded assumptions, is because of a felicitous accident of
Gaussian statistics rather than for reasons of thermodynamics and
statistical mechanics.

\section{Discussion}

\noindent 1. As said before, it is clearly impossible to do
justice to all the publications that have appeared in connection
with the JE. We have confined ourselves here to represent the
contents of two articles on the {\it{general foundations}} of the
JE on Statistical Mechanics by
Jarzinski$^{[1,2]}$.\\
2. In addition we discussed an important experiment which appears
to confirm the JE, since the system is in aqueous solution, close
to equilibrium and the processes are (quasi) quasi-static and
therefore nearly reversible, in spite of some hysteresis.\\
3. Although in Liphardts' experiment clear hysteresis loops can be
seen, they occur over relatively small parts of the trajectories,
which mostly appear to be reversible. True irreversible processes,
where e.g. no temperature can be defined in the system, have so
far not been considered experimentally (because of the liquid
surroundings used so far). Such irreversible processes would
question severely the assignments of a canonical weights
$e^{-\beta W_{irr}}$, with $\beta$ the temperature of the heat
bath.

We note that although to observe the JE the systems have to be
small, there is a gap between the experiments performed on
biochemical systems (in aqueous solutions) and those on small
physical (or chemical) systems without an aqueous heat bath, which
have not yet been performed.\\
4. There are many different theoretical {\it{models}}, to mimic
the heat bath, which have been used to derive the JE. They are all
stochastic and, as far as we are aware, all use Markovian and
detailed balance properties, which place them, in our opinion,
near predominantly reversible heat exchanges. We confine ourselves
again to two examples.

a) This holds for Crooks'$^{[8]}$ derivation of the JE for a
simple Ising model. He discretizes there the phase space
trajectory in sequences of two sub steps, whereby first a control
parameter $\lambda$ does instantaneous work $W$ on the system
followed by an exchange of thermal energy $E$ with a heat bath.
Either the first forward work step is followed by a heat exchange
step which lasts sufficiently long compared to the characteristic
relaxation times of the system that a state of canonical
equilibrium is reached (cf.Crooks' eq.(8)) in which case the
process is a discrete sequence of equilibrium states and therefore
a reversible process (like in a reversible experiment) and the JE
will be obtained as that for a reversible process. Or the time
after the first forward step  is not sufficiently long to obtain
an equilibrium state, in which case an irreversible process takes
place, which may be close to a reversible one, but no detailed
balance obtains at any $\lambda_i$.{\footnote{The assumed detailed
balance at every step $\lambda_i$ is not a direct consequence of
microscopic reversibility alone, and has been proved so far only
for thermal equilibrium [13,14].}}

Similarly Jarzynski's procedure applied to the Langevin
Equation$^{[2]}$, because of the $e^{-\beta W_{irr}}$ weight
assumption, is restricted to systems near equilibrium, with near
reversible heat exchanges.

b) Jarzynski made numerical calculations$^{[9]}$ of which we will
discuss only that on a harmonic oscillator.  He considered $10^5$
simulations of the work done on a single harmonic oscillator,
whose frequency is switched from $\omega_0 =1.0$ to $\omega_1 =
2.0$ over a switching time $t_s$. The Hamiltonian of the harmonic
oscillator is:
\begin{equation}
H = \frac{p^2}{2} + \omega^2_\lambda \; \frac{x^2}{2}
\end{equation}
where $\omega_\lambda$ switches from $\omega_0$ to $\omega_1 = 2
\omega_0$. If the change from $\omega_0$ to $\omega_1$ proceeds
infinitely slowly ($t_s \rightarrow \infty$) and adiabatic
invariance can be applied, i.e. $H/\omega$ = constant, then
Jarzynski obtains, for a canonical distribution of initial
energies a distribution function for the work $W$. Using this to
compute the average of the exponential work $e^{-\beta W}$, gives
$\beta^{-1} \ln (\omega_1/\omega_0) = 0.693 \beta^{-1} = \Delta
F$, since $\omega_1/\omega_0$ is just the ratio of the canonical
partition functions (cf.Jarzynski in ref.9 eq.(59) and fig.3
($W^x$ points)).  One can get an idea of the average value of
$\beta W$ for fast switchings ($t_s=1$), by using a similar
calculation.

One finds then for the arithmetic average of $\beta W$ the result
$(\frac{\omega_1}{\omega_0}- 1)= 1$. Adding then $W_{rev} = 1$, a
value of 2 is obtained not far from the numerical result (ref.7,
fig.3 ($W^a$). However, this result depends critically on the
assumption of adiabatic invariance for $t_s=1$, while this
actually only holds for $t_s \rightarrow \infty$. Moreover, if for
example, the energy would change as $\omega^2$, the Jarzynski
weighting would give $2\Delta F$ and the arithmetic weighting
would give a value 4 instead of 2 for $\beta W$.

Jarzynski claims that the JE holds for all systems (not only
harmonic oscillators), while we see that the JE is critically
dependent on adiabatic invariance for finite $t_s$, even for the
harmonic oscillator.\\
5. Finally we note that eq.(1) implies that a remarkable new
relation would follow from the $JE$, if it were an exact equality,
viz.:
\begin{equation}
<e^{-\beta W_{irr}}> =  <e^{-\beta W_{rev}}>
\end{equation}
In the literature, the eq.(6), or equivalently $<e^{-\beta W_d}> =
1$, is justified by invoking an apparently very fortuitous general
cancellation due to the weighting of $W_d>0$ and $W_d<0$ with
$e^{-\beta W_d}$. It is entirely unclear to us, how this can be
achieved for irreversible processes in a system, for all
$\lambda(t)$ and all corresponding phase space paths, using in all
cases the (unconnected) heat bath temperature $\beta$ in the
weights.

This would imply the equality of the average over the
exponentiated work of all irreversible paths from $A$ to $B$, to
the (average over all the) reversible path(s) from $A$ to $B$.
Considering the unknown nature and wide variety of all
irreversible paths this equality does not seem physically
understandable. If true, it would incorporate a hitherto unknown
symmetry for irreversible processes for any switching rate, i.e. a
new extension of the Second Law. To be sure there are special
cases, like those treated in section 4, sub 2 and in section 5,
sub 4b, where this equality holds under certain specific
assumptions for irreversible processes near equilibrium.
\newpage

\noindent{\bf{References}}

\noindent 1. C. Jarzynski, {\it{Phys.Rev.Lett.}} {\bf{78}}, 2690
(1997).\\
2. C. Jarzynski, in {\it{Dynamics of Dissipation}}, P.
Garbaczewski, R. Olkiewicz, eds., (Springer, Berlin 2002) p.63
and references therein.\\
3. F. Ritort, {\underline{S\'{e}minaire Poincar\'{e}}}, 6/12/2003,
p.63.\\
4. J. Liphardt, S. Dumont, S. B. Smith, I. Tinoco, Jr., and C.
Bustamante, {\it{Science}} {\bf{296}}, 1832 (2002).\\
5. L. D. Landau and E. M. Lifshitz, {\it{Statistical Physics}},
Addison Wesley, Reading, Mass. (1958) p.45, Eq.(15.1).\\
6. S. K. Blau, {\it{Physics Today}}, September (2002) p.19.\\
7. W. Zhang and S-J. Chen, {\it{Proc. Natl. Acad. Sci.USA}},
{\bf{99}}, 1931 (2002).\\
8. G. E. Crooks, {\it{J. Stat. Phys.}} {\underline{90}}, 1481 (1998).\\
9. C. Jarzynski, {\it{Phys. Rev. E}} {\underline{56}}, 5018
(1997).\\
10. Ref. 4, Supplementary Material.\\
11. Ritort et al, PNAS {\underline{99}}, 13544 (2002).\\
12. Gore et al, PNAS, {\underline{100}}. 12564 (2003).\\
13. S. R. de Groot and P. Mazur, {\it{Non-Equilibrium
Thermodynamics}}, Dover (1984) p.93.\\
14. N. G. van Kampen, {\it{Stochastic Processes in Physics and
Chemistry}}, Elseveier, Amsterdam, V.6 (1992).\\

\noindent {\underline{Acknowledgement}}

\noindent The authors are indebted to Professors R. F. Fox, J. M.
Kincaid, and B. Widom and Drs. R. van Zon, E. van Nimwegen and T.
Tuschl for helpful discussions. EGDC also gratefully acknowledges
support from the Office of Basic Energy Sciences of the US
Department of Energy under Grant number DE-FG02-88-ER13847.\\

\end{document}